\documentclass[prb,showpacs,preprint,superscriptaddress]{revtex4}
\usepackage{amssymb,graphicx,amsbsy}

\begin{document}

\title{Quantum Monte Carlo study of the transverse-field quantum Ising model
on infinite-dimensional structures}
\author{Seung Ki Baek}
\affiliation{Department of Physics, Sungkyunkwan University, Suwon 440-746,
Korea}
\affiliation{Integrated Science Laboratory, Ume{\aa} University, S-901 87
Ume{\aa}, Sweden}
\author{Jaegon Um}
\affiliation{School of Physics, Korea Institute of Advanced Study, Seoul
130-722, Korea}
\author{Su Do Yi}
\affiliation{BK21 Physics Research Division and Department of Physics,
Sungkyunkwan University, Suwon 440-746, Korea}
\author{Beom Jun Kim}
\email[Corresponding author, E-mail: ]{beomjun@skku.edu}
\affiliation{BK21 Physics Research Division and Department of Physics,
Sungkyunkwan University, Suwon 440-746, Korea}

\begin{abstract}
In a number of classical statistical-physical models, there exists a
characteristic dimensionality called the upper critical dimension above
which one observes the mean-field critical behavior. Instead of constructing
high-dimensional lattices, however, one can also consider
infinite-dimensional structures, and the question is whether this mean-field
character extends to quantum-mechanical cases as well. We therefore
investigate the transverse-field quantum Ising model on the globally coupled
network and on the Watts-Strogatz small-world network by means of quantum Monte
Carlo simulations and the finite-size scaling analysis. We confirm that both
of the structures exhibit critical behavior consistent with the mean-field
description. In particular, we show that the existing cumulant method
has difficulty in estimating the correct dynamic critical exponent and
suggest that an order parameter based on the quantum-mechanical
expectation value can be a practically useful numerical observable to
determine critical behavior when there is no well-defined dimensionality.
\end{abstract}

%05.30.Rt	quantum phase transitions
%75.10.Jm	quantized spin models
%89.75.Hc 	Networks and genealogical trees 
\maketitle

The effects of geometry on physical properties have drawn attention from various
areas of statistical physics. First of all, we note the extensive studies on
complex networks (see Ref.~\onlinecite{doro} for a review). Aside from their
ubiquity and practical importance, a remarkable point is that many of the
complex networks tend to enhance correlations dramatically when a
statistical-physical system is put on top of them. Let us look at the
Watts-Strogatz (WS) network,~\cite{ws} for example: it starts from a
one-dimensional (1D) ring of size $N$ where each point is linked to its $2k$
nearest neighbors. Among the total $Nk$ bonds, we pick up $PNk$ bonds with
$0<P<1$, and then for each of them, we change one of its endpoints to a
randomly chosen site. The WS network is a classical model of a small-world
network characterized by two structural features: the length scale $L$ is
only logarithmic with respect to the system size $N$ and the clustering
coefficient is nevertheless relatively high.~\cite{ws} Due to the
small-world property, one observes the mean-field critical behavior in many
cases, such as percolation, Ising, and $XY$ spin
models.~\cite{moore,smallxy,hong,med} Since that is expected only for
very-high-dimensional structures, many complex networks, including the WS
network, are often called infinite dimensional. One may notice that this
terminology also makes sense in that $L \sim N^{1/d}$ in $d$-dimensional
lattices while the WS network has only $L \sim \log N$. The theory of
finite-size scaling has been well developed for such complex networks
in combination with extensive numerical calculations.~\cite{ref}

Compared to the recent progress in the study of classical systems in these
structures, however, there is relatively little known about quantum cases.
Even the transverse-field quantum Ising system on the Bethe lattice has been
examined very recently, and it is still to be clarified whether the phase
transition is really of the mean-field type,~\cite{nagaj,krz} and the same
model on the WS network has been checked only in terms of the thermal phase
transitions.~\cite{yi} In this work, we thus investigate how the quantum
Ising spin system in the transverse field behaves in the infinite-dimensional
globally coupled network and the WS network.
We do not consider the scale-free network in this work since the
critical behavior can be possibly far richer than the conventional mean-field
description.~\cite{yi3,yi2} The transverse-field Ising model is defined by
the quantum-mechanical Hamiltonian
\begin{equation}
H = -J \sum_{\left< ij \right>} \sigma^z_i \sigma^z_j + \Gamma \sum_i
\sigma^x_i,
\label{eq:ham}
\end{equation}
where ${\vec \sigma}_i \equiv (\sigma_i^x, \sigma_i^y, \sigma_i^z)$
is a spin-$\frac{1}{2}$ operator at each given site $i$ and
the summation runs over all the nearest-neighbor
pairs. This Hamiltonian also contains a constant $J>0$ to represent the
ferromagnetic-interaction strength and another constant $\Gamma$ to represent
the transverse-field strength. If located in $d$ dimensions,
this system exhibits a phase transition at a certain critical-field strength,
$\Gamma = \Gamma_c$, and its criticality is known to be equivalent to that
of the classical $(d+1)$-dimensional Ising model where $\Gamma$ is replaced
by temperature $T$.~\cite{elliott,pfeuty,um} But it is not entirely obvious
what is going to happen when the dimensionality lacks its precise meaning
in the infinite-dimensional structures, such as the WS network.
Before studying the WS network, however,
it would be good to check another infinite-dimensional structure, i.e.,
the globally connected network, since the transverse-field Ising system on
this structure has been already known to exhibit the zero-temperature quantum
phase transition of the mean-field universality by the Hamiltonian-matrix
diagonalization.~\cite{botet} The globally coupled network is simply
obtained by connecting every site to all of the others, but the Hamiltonian
needs a little modification to Eq.~(\ref{eq:ham}) in order to make it
extensive as follows:
\begin{equation}
H = -N^{-1}J \sum_{i \neq j} \sigma^z_i \sigma^z_j + \Gamma \sum_i
\sigma^x_i.
\label{eq:global}
\end{equation}
The main tool in analyzing Eqs.~(\ref{eq:ham}) and (\ref{eq:global}) is the
world-line quantum Monte Carlo method,~\cite{kawa2}
where the system is endowed with an additional temporal dimension in the
imaginary-time direction. The length in this direction is the same as the
inverse temperature $b \equiv 1/T$ and divided into $N_t$ slices. The
Suzuki-Trotter decomposition says that we get correct results of the
original transverse-field Ising model at temperature $T$ as $N_t \rightarrow
\infty$.
Our quantum Monte Carlo algorithm is
based on the single-cluster flip algorithm~\cite{wolff} and employs the idea of
continuous imaginary time, as described in Ref.~\onlinecite{kawa}. For
convenience, we actually work with integer variables in programming (instead
of floating variables in indexing) the imaginary-time axis, and then make the
range of the indices, $N_t$, up to $O(10^8)$. This virtually eliminates the
error involved in the width of an imaginary-time slice in the Suzuki-Trotter
decomposition. We have checked that the implemented algorithm convincingly
reproduces the expected results for $d=1$ and $2$ where the dynamic critical
exponent is already known to be $z=1$.~\cite{sachdev} The dynamic critical
exponent $z$ tells us how to scale the temporal size $b$ with respect to $L$
to observe correct critical behavior, so it means that $b \sim L^z = L^1$ in
ordinary $d$-dimensional lattices. Note that this quantity is required since
our quantum Monte Carlo algorithm cannot directly access $T=0$, but needs a
suitable extrapolation of $T$ in relation to the system size, unless one
simulates extremely low $T$ compared to all of the finite sizes under
consideration. When $z$ is known {\it a priori}, everything is just
straightforward. Applying the algorithm to the infinite-dimensional
structures described above, however, we usually do not know it at the
starting point and need to estimate $z'$ to scale $T \sim N^{-z'}$ in a
suitable way. We can formulate the problem as follows: let us consider the
magnetic order parameter $|m|$ taken from the whole $(d+1)$-dimensional
space-time region, that is
\begin{equation}
\left< |m| \right> = N^{-1}_{t} N^{-1} 
\left<  \left | \sum_{i,t} \sigma^{z}_{i,t} \right | \right>,
\label{def:m}
\end{equation}
where each time slice is indexed by $t$ and the bracket $\left< \cdots
\right>$ means the thermal average, as well as the disorder average in the case
of the WS network. Note that $\sigma_{i,t}^z$ in Eq.~(\ref{def:m}) is not an
operator, but a spin value projected onto the $z$ direction at a certain
space-time. One can argue a finite-size scaling ansatz in the following
two-parameter form:
\begin{equation}
\left< |m| \right> = N^{-\beta/\nu'} f \left[ (\Gamma - \Gamma_c)
N^{1/\nu'}, T N^{z'} \right].
\label{eq:fss}
\end{equation}
It is usually complicated to directly deal with this two-parameter scaling
form. But when $T$ is suitably scaled with a dynamic critical exponent $z'$
so that $\tau \equiv T N^{z'} = {\rm const.}$, Eq.~(\ref{eq:fss}) reduces to
a usual scaling form with a single parameter, from which it is possible to
determine the critical exponents. We therefore follow the procedure proposed
in Refs.~\onlinecite{guo} and \onlinecite{young} to estimate $z'$. The idea
is that Binder's cumulant for a finite-sized system,
\begin{equation}
U = 1 - \frac{\left< m^4 \right>}{3 \left< m^2\right>^2},
\label{eq:bin}
\end{equation}
vanishes not only in the limit of $T \rightarrow \infty$, but also in the
limit of $T \rightarrow 0$ where the system virtually becomes a classical 1D
Ising chain elongated in the imaginary-time direction. One may thus expect
that a maximum will appear by $U = U_{\rm max}$ at a certain characteristic
temperature $T_{\rm max}$ for each given system size $N$.
\begin{figure}
\includegraphics[width=0.48\textwidth]{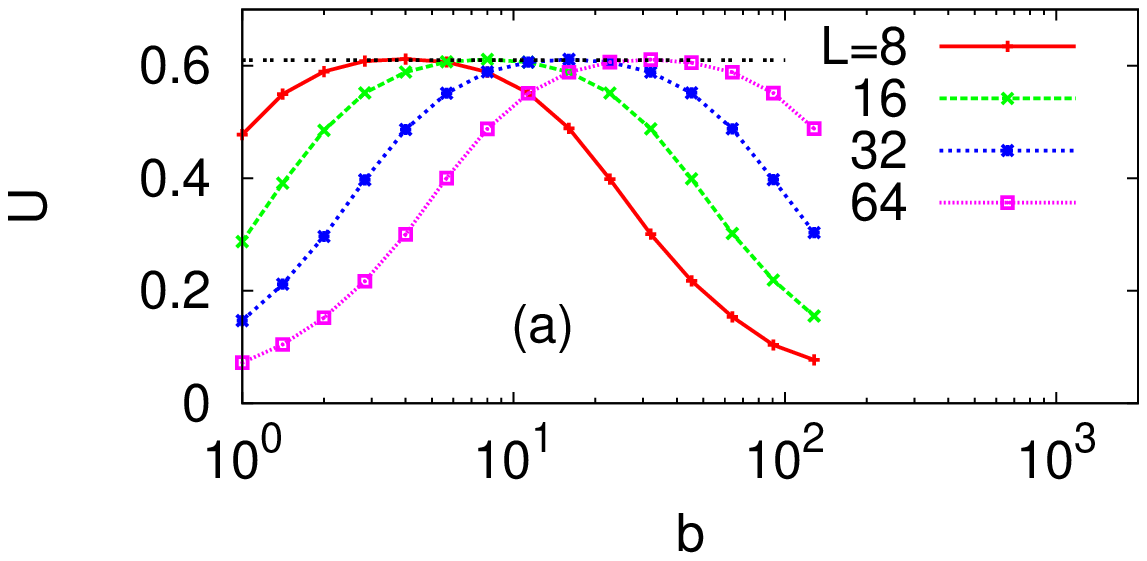}
\includegraphics[width=0.48\textwidth]{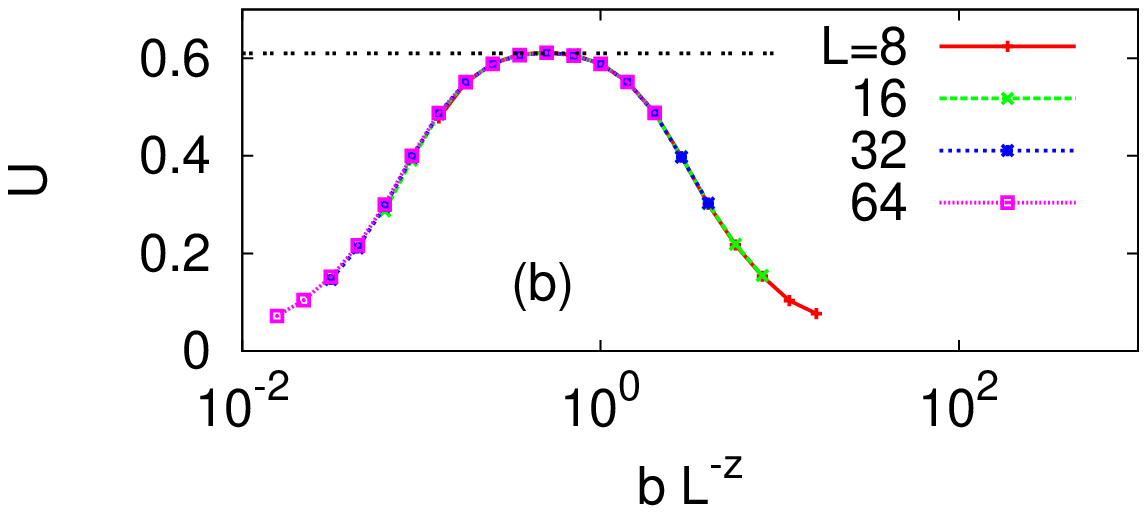}
\caption{(Color online)
(a) Binder's cumulant for the 1D transverse-field Ising model at
$\Gamma=\Gamma^{\rm (1D)}_c=1$ as a function of the inverse temperature $b =
1/T$. The transverse-field strength $\Gamma$ is in units of $J$ and the
temperature $T$ is in units of $J/k_B$ where $k_B$ is the Boltzmann
constant. (b) Scaling collapse with the dynamic critical exponent $z=1$. The
dotted lines indicate its universal amplitude ratio for the two-dimensional
Ising model [$U^* = 0.610692(2)$]. }
\label{fig:1d}
\end{figure}
If the field strength is above the critical threshold by $\Gamma >
\Gamma_c$, then the peak value $U_{\rm max}$ will eventually vanish as $N$
becomes larger, but will approach a nonzero value if $\Gamma < \Gamma_c$
instead. This speculation provides a way to determine $\Gamma_c$ where
$U_{\rm max}$ remains constant regardless of $N$. We illustrate how it looks
for the 1D case at $\Gamma_c^{\rm (1D)} = 1$ in Fig.~\ref{fig:1d}(a). Note
the nice symmetry as the size and inverse temperature $b$ vary at the same
time. It implies that one can also assume that $U$ obeys a finite-size
scaling form as
\begin{equation}
U = g \left[ (\Gamma - \Gamma_c) N^{1/\nu'}, T N^{z'} \right].
\label{eq:ufss}
\end{equation}
Therefore, $U$ is a function of $\tau = TN^{z'}$ at $\Gamma = \Gamma_c$, and
a correct $z'$ will yield a scaling collapse of $U$ [Fig.~\ref{fig:1d}(b)].
It is also notable that $U_{\max}$ in Fig.~\ref{fig:1d} is fully consistent
with the universal amplitude ratio $U^{\ast} = 0.610692(2)$ for the
two-dimensional Ising model.~\cite{salas} By checking the size dependence of
$b_{\max} \equiv 1/T_{\max}$ where $U$ reaches the maximum, one will be able
to obtain the dynamic critical exponent. Then, we calculate the magnetic
order parameter with changing both $N$ and $T$, and finally get the critical
exponents in the zero-temperature thermodynamic limit by the standard
finite-size scaling, $\left<|m|\right> \sim N^{-\beta/\nu'}$ and $dU/d\Gamma
\sim N^{1/\nu'}$ at $\Gamma = \Gamma_c$.

In addition, it is interesting to note that in Fig.~\ref{fig:1d}(b)
the curve exhibits reflection symmetry with the peak position as the axis of
symmetry. The reason for this
can be argued in the following way: for given $L$, let
us begin with observing the system at $b = b_{\rm max}$. Then one can say
that the correlation length in the spatial direction amounts to $L$, while
that in the temporal direction amounts to $b$. Let us increase the
temperature so that $b$ becomes $b_{\rm max}/2$. When a system has different
lengths in different axes, the correlation length will be bounded by the
shortest one, so the correlation length will become about one half compared
to the previous case of $b=b_{\rm max}$. It means that as we traverse the
system in the spatial direction, we will find no correlation at a distance
around $L/2$, and we may roughly regard the system as separated into two
uncorrelated parts in the spatial direction. If we instead lower the
temperature to $b = 2 b_{\rm max}$, the correlation length still remains
bounded by $L$, and we will find no correlation as we traverse the system by
$b_{\rm max}$ in the temporal direction. In this sense, we may regard the
system as separated into two uncorrelated parts in the {\em temporal}
direction this time. It becomes precisely symmetric to the case of $b =
b_{\rm max}/2$ by exchanging the axes and adjusting the overall scale. In
other words, there is symmetry between the cases of $\log b_{\rm max}
\rightarrow \log b_{\rm max} \pm \log 2$, which explains the reflection
symmetry in Fig.~\ref{fig:1d}(b). If we extend this argument to a higher
dimension, say $d>1$, the system will be broken into $2^d$ pieces when we
take $b \rightarrow b_{\rm max}/2$. Since the other part of the argument
about $b \rightarrow 2 b_{\rm max}$ remains the same, it predicts that $U$
will decay faster on the low-$b$ side than on the high-$b$ side, and that
this asymmetry will be more pronounced as $d$ gets higher.

\begin{figure}
\includegraphics[width=0.48\textwidth]{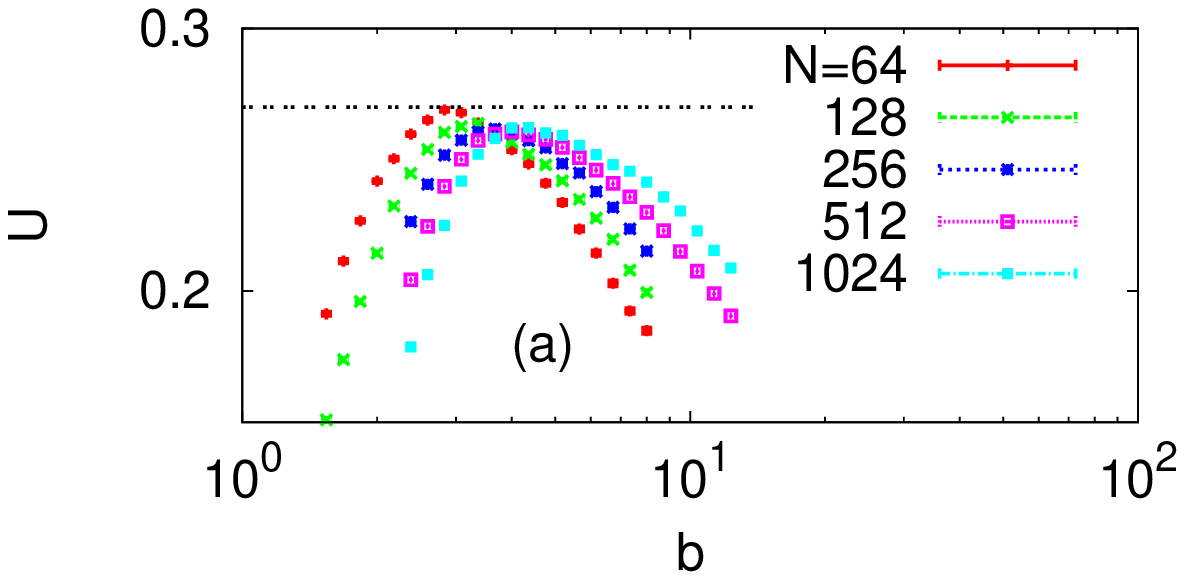}
\includegraphics[width=0.48\textwidth]{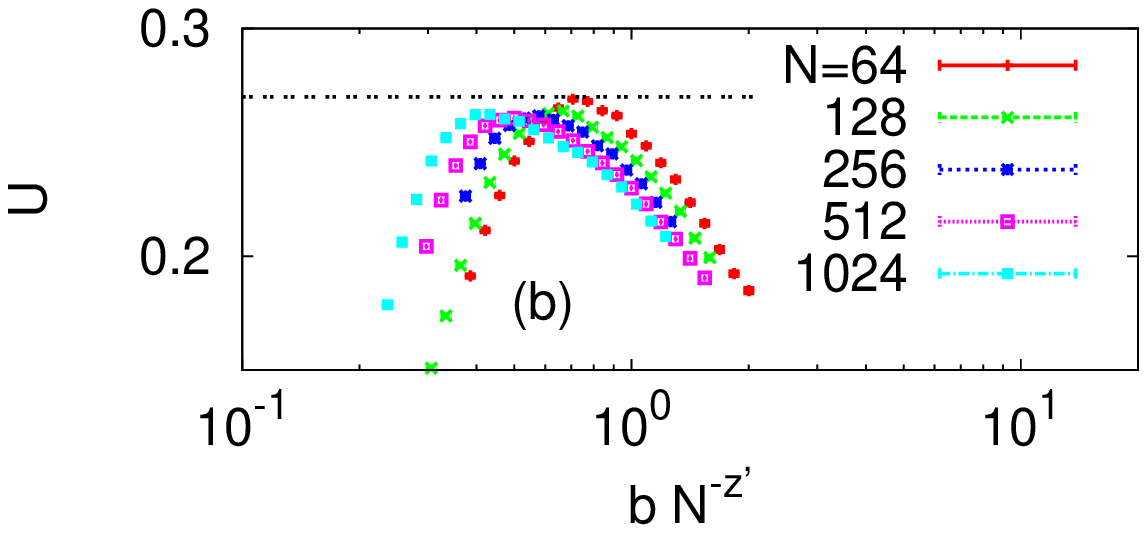}
\includegraphics[width=0.48\textwidth]{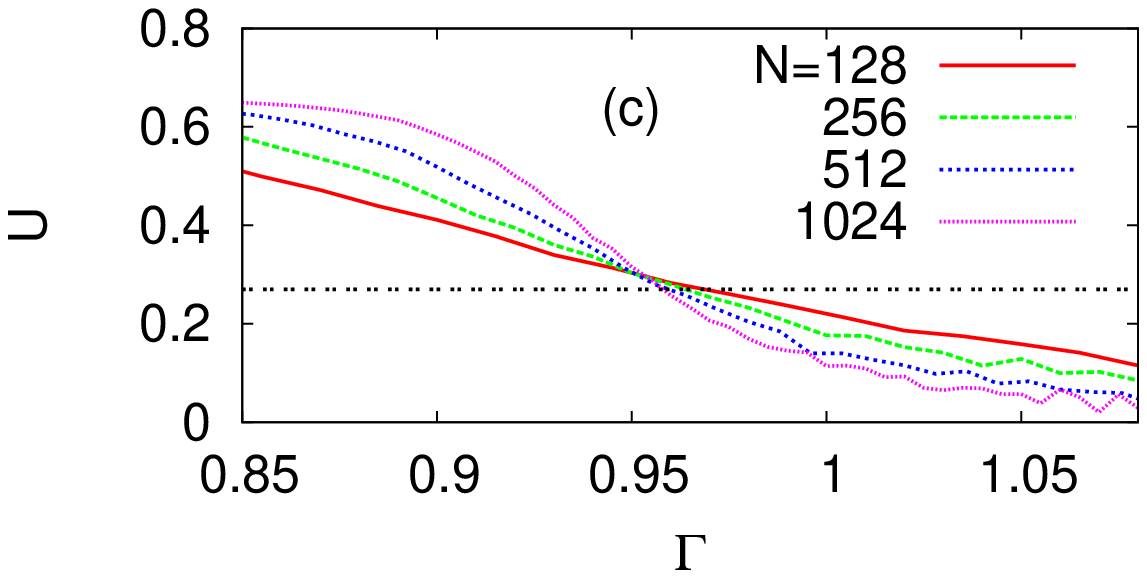}
\includegraphics[width=0.48\textwidth]{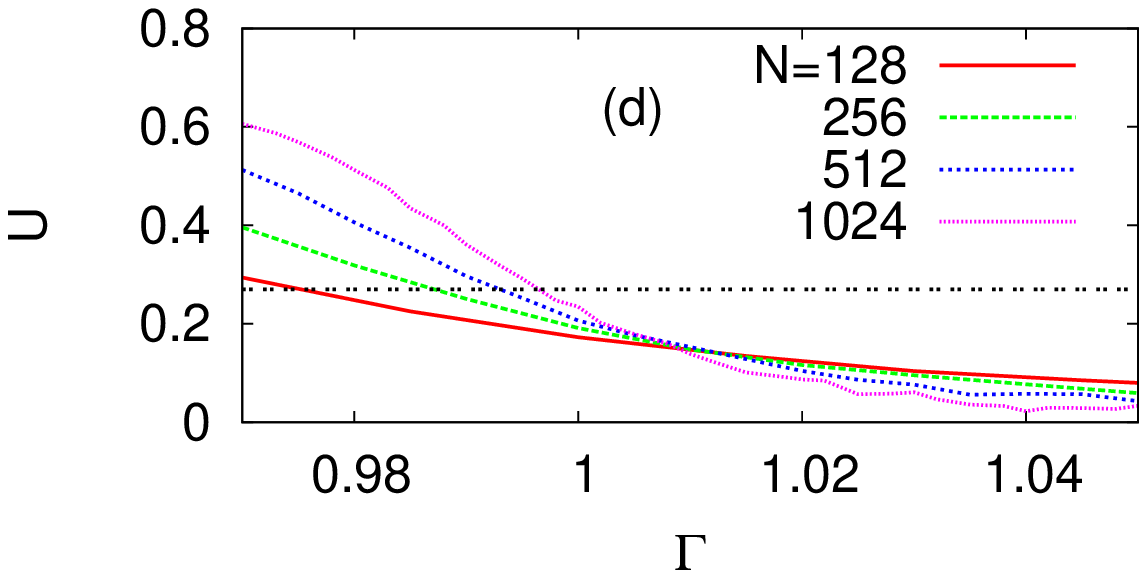}
\includegraphics[width=0.48\textwidth]{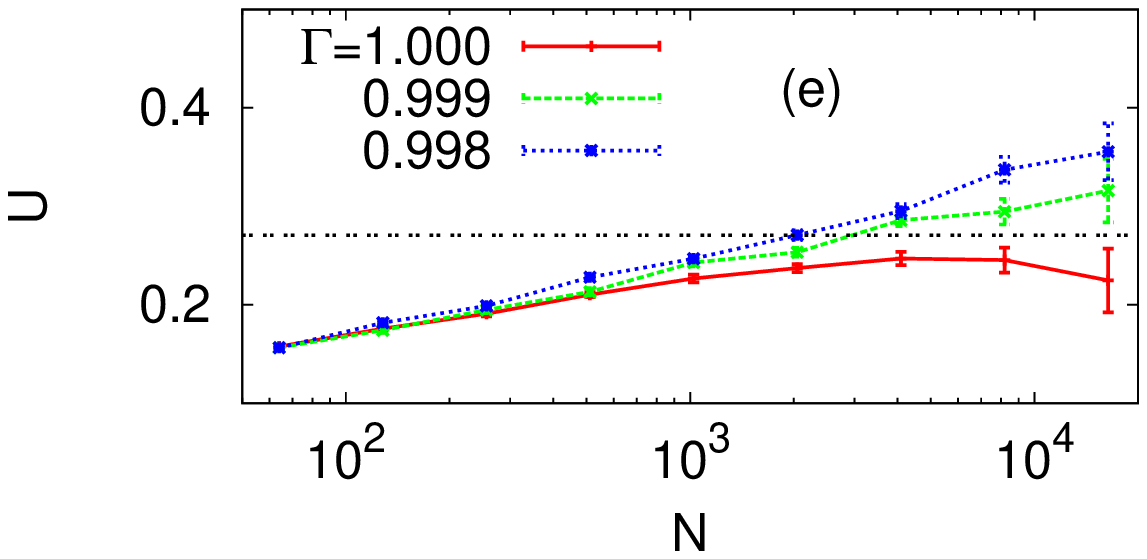}
\includegraphics[width=0.48\textwidth]{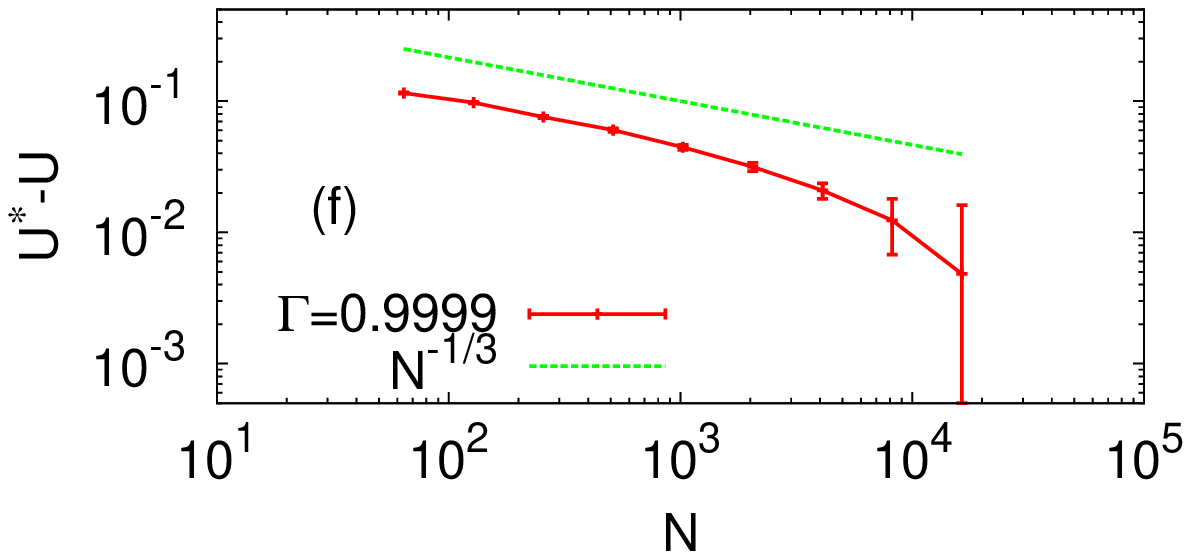}
\caption{(Color online) Observations for the globally coupled network. (a)
Binder's cumulant at $\Gamma=\Gamma_c^{\rm (G)}=1$. (b) Scaling of the
horizontal axis with the known dynamic critical exponent $z'=1/3$ [compare
with Fig.~\ref{fig:1d}(b)]. (c) Binder's cumulant at $T=0.5$ shows a
crossing point, already close to the universal value, while (d) the crossing
is still moving due to finite-size effects at $T=0.1$.
(e) Binder's cumulant as a function of $N$ at three different $\Gamma$'s,
with temperature $T=0.1$ as in (d).
The dotted lines
represent the universal amplitude ratio for the mean-field case [$U^{\ast}
\approx 0.270520$].
(f) The difference of $U$ from $U^{\ast}$ at $\Gamma = 0.9999$, in
comparison with $N^{-1/3}$.
 }
\label{fig:global}
\end{figure}

Let us check how the method using Eq.~(\ref{eq:ufss}) works for the globally
coupled network with Eq.~(\ref{eq:global}). As mentioned above, it has been
already established that the transverse-field quantum Ising model on it
exhibits the mean-field critical behavior, and the critical threshold is at
$\Gamma_c^{\rm (G)} = 1$ in units of $J$.~\cite{botet} Since the
transverse-field Ising model has the upper critical dimension $d_u = 3$, it
also leads to $z' = z/d_u = 1/3$.~\cite{botet} We thus calculate Binder's
cumulant at $\Gamma=1$ and depict the results in Fig.~\ref{fig:global}(a).
We immediately notice two features from the figure: first, the maximum
height is close to the universal amplitude ratio for the mean-field model,
$U^{\ast} \approx 0.270520$.~\cite{lb} Second, the curve shapes become more
asymmetric as $N$ gets larger. This observation is qualitatively explained
by the argument in the paragraph above, and furthermore implies that it is
not possible to make a scaling collapse of this plot just by tuning $T$ in
accordance with $N$ as postulated in Eq.~(\ref{eq:ufss}), at least up to the
sizes that we have in this work [Fig.~\ref{fig:global}(b)]. It hinders one
from using the extrapolation to $N \rightarrow \infty$ and $T \rightarrow 0$
from the finite-sized data at finite temperatures by Eq.~(\ref{eq:ufss}).
Even in the case that one works at a fixed temperature, the finite-size
effects may well enter and dominate the situation: when the temperature is
relatively high, say $T=0.5$, we have little difficulty finding the
mean-field transition at $\Gamma_c \approx 0.958(5)$ with the {\it
classical} upper critical dimension $d_u^{\rm (classic)} = 4$, and the value
of $U$ is close to the universal amplitude ratio [Fig.~\ref{fig:global}(c)
and its scaling collapse (not shown here)].
This observation can be explained as follows:
rewriting Eq.~(\ref{eq:ufss}) as $U = g(TN^{z'})$ at
$\Gamma=\Gamma_c$, one may expect $g$ to converge to a constant when the
system gets large enough, i.e., $N \gg T^{-1/z'}$. It in turn means that the
scaling form in Eq.~(\ref{eq:ufss}) becomes insensitive to the second
argument at a sufficiently high $T$. This is why we find such finite-size
scaling behavior with varying $\Gamma$ at $T=0.5$ in
Fig.~\ref{fig:global}(c). The classical mean-field behavior at this $T$
clearly confirms a consequence of crossover phenomena from the quantum to
the classical critical behavior at different temperatures.~\cite{cross}
However, if we move to a low but still finite temperature, i.e., $T=0.1$ in
this example, the finite-size effect becomes so substantial that it gets hard
to characterize the transition [Fig.~\ref{fig:global}(d)]. Note that the
crossing point is still moving to the low-$\Gamma$ side, rising up very
slowly to $U^{\ast}$.
If we increase $N$ by more than an order of magnitude,
Binder's cumulant begins to move in the correct direction
[Fig.~\ref{fig:global}(e)]. The critical-field strength at $T=0.1$ can be
estimated to differ from $\Gamma=1$ by less than $5 \times 10^{-4}$, and the
cumulant will behave as $U = U^{\ast} (1 + c N^{-\Delta} + \cdots)$ with a
coefficient $c$ and a correction exponent $\Delta \ge 0$ at this critical
point.
Since Eq.~(\ref{eq:ufss}) leads to $U = g( bN^{-z'})$
at $\Gamma = \Gamma_c$, one may guess that $\Delta = z' = 1/3$ by expanding
the scaling function $g$ into a Taylor series. This simple guess is not very
far from our numerical observation shown in Fig.~\ref{fig:global}(f).
To sum up, our results suggest that Eq.~(\ref{eq:bin}) might not be a
very efficient numerical observable to reproduce the correct results for
infinite-dimensional structures.

\begin{figure}
\includegraphics[width=0.48\textwidth]{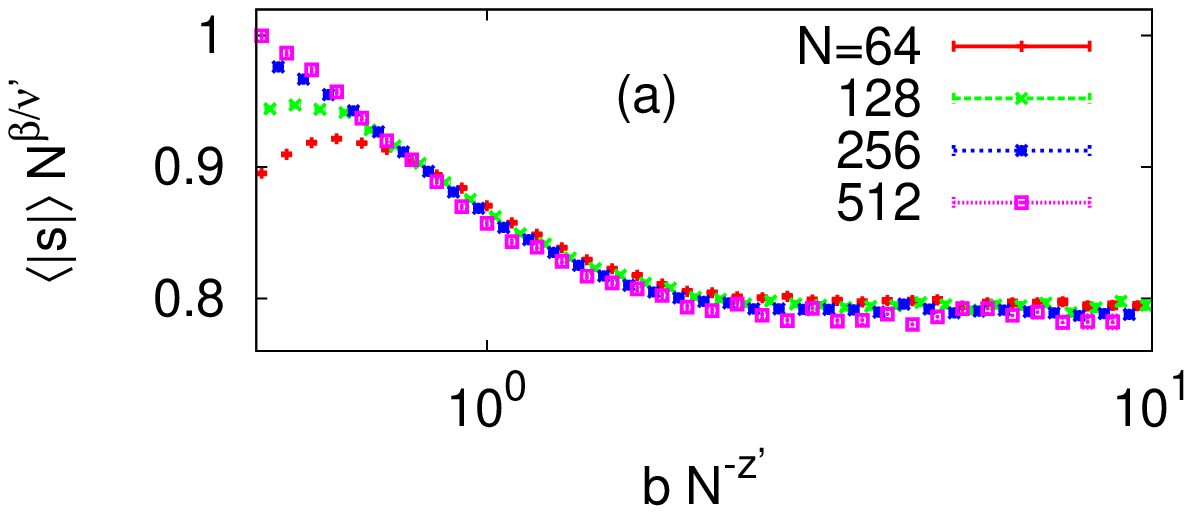}
\includegraphics[width=0.48\textwidth]{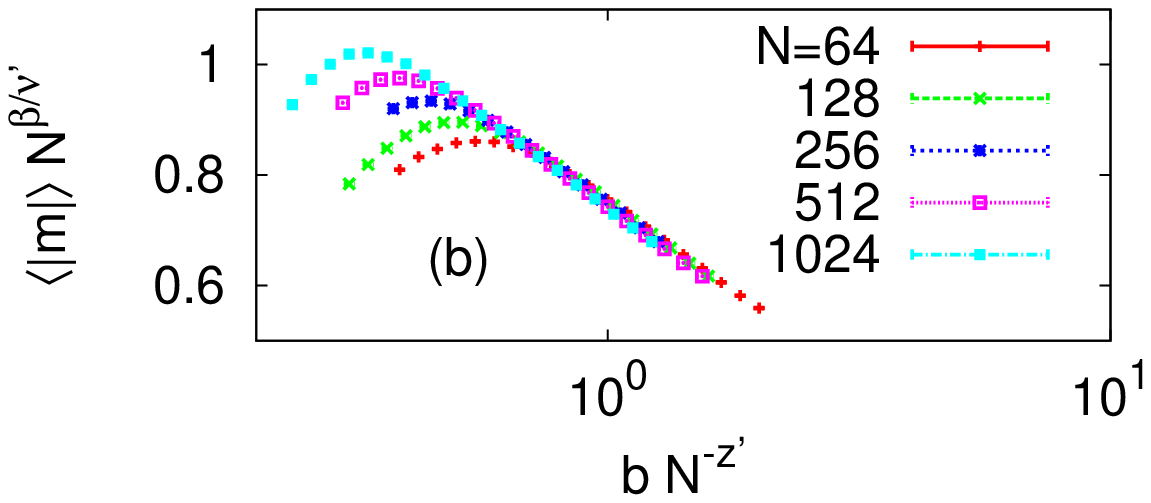}
\includegraphics[width=0.48\textwidth]{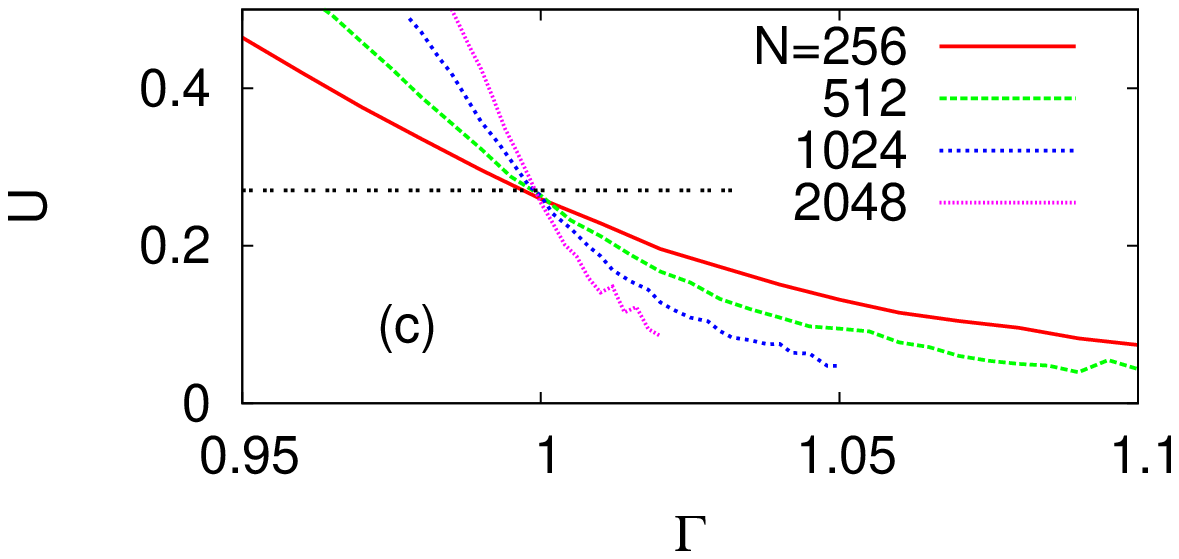}
\includegraphics[width=0.48\textwidth]{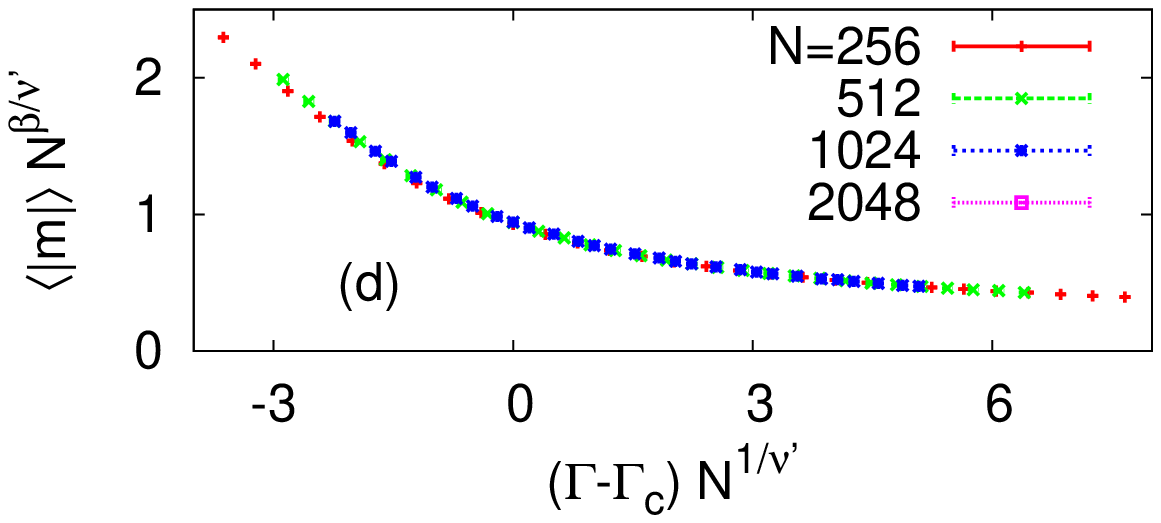}
\caption{(Color online) Observations for the globally coupled network. (a)
Magnetization at a fixed imaginary time [Eq.~(\ref{def:s})] yields scaling
collapse with $z'=1/3$. (b) Scaling of the magnetic order parameter
[Eq.~(\ref{def:m})] with $z'=1/3$ at $\Gamma = 1$. By scaling $T$ with
this $z'$ in such a way that $\tau = TN^{z'}=2$, (c)
Binder's cumulant and (d) the magnetic order parameter show the correct
mean-field quantum phase transition at $\Gamma_c^{\rm (G)}=1$ with
$\beta/\nu' = 1/3$ and $1/\nu'=2/3$. The dotted line represents the
universal amplitude ratio for the mean-field case [$U^{\ast} \approx
0.270520$]. }
\label{fig:global2}
\end{figure}

As discussed above, although intuitively appealing, the concept of the
characteristic temperature $T_{\rm max}$ where $U$ shows a maximum is not
adequate for high-dimensional structures. The difficulty can be overcome by
using an alternative quantity to $U$. Let us take the magnitude of
spontaneous magnetization at an arbitrary {\em fixed} point in the
imaginary-time axis,
\begin{equation}
\left< |s| \right> = N^{-1} \left< \left| \sum_{i} \sigma^{z}_{i,t}
\right | \right>.
\label{def:s}
\end{equation}
Due to the time translational symmetry, $\left< |s| \right> $ defined in
Eq.~(\ref{def:s}) is a time-independent value, and, moreover, it corresponds
to the Boltzmann statistics with eigenfunctions diagonalizing the
Hamiltonian Eq.~(\ref{eq:ham}) or Eq.~(\ref{eq:global}), i.e.,
\[ \left< |s| \right> = Z^{-1} \sum_{j}
|s_j| e^{-b E_{j}} \]
where $Z$ is the partition function, and $|s_j|\equiv \sum_{l} \left|
a_{l}^{j} \right|^{2} \left| \left< s_{l} \right|  \sum_{i} \sigma^{z}_{i}
\left|s_{l} \right> \right|$ and $E_{j}$ are the quantum-mechanical
expectation value of the operator $\sum_i \sigma_i^z$ and the
energy eigenvalue for the $j$th eigenstate $\left|\Psi_{j} \right> \equiv
\sum_{l} a_{l}^{j} \left| s_{l} \right>$, respectively. Consequently, as $T
\rightarrow 0$, $\left< |s| \right>$ goes to a nonzero expectation value of
the ground state for finite $N$, which is different from the behavior of
$\left< |m| \right>$ in Eq.~(\ref{def:m}). The scaling form is again
considered as
\begin{equation}
\left< |s| \right> = N^{-\beta/\nu'} h \left[ (\Gamma - \Gamma_c)
N^{1/\nu'}, T N^{z'} \right],
\label{eq:sfss}
\end{equation}
so one finds scaling collapse $\left< |s| \right> N^{\beta/\nu'} = h ( T
N^{z'} )$ at $\Gamma=\Gamma_c$. Furthermore, as mentioned above,
Fig.~\ref{fig:global2}(a) shows that this quantity actually converges to a
constant at $b \rightarrow \infty$ for each $N$. Since $\left< |s| \right>$
is insensitive to $T$ as $T \rightarrow 0$, we can drop out the second
argument of the scaling function $h$ in Eq.~(\ref{eq:sfss}) at sufficiently
low $T$ and we are back to the single-parameter scaling form,
\[ \left< |s| \right> \approx N^{-\beta/\nu'} h \left[ (\Gamma - \Gamma_c)
N^{1/\nu'} \right] \]
around $\Gamma_c$. Fitting the values of $\left< |s| \right>$ at $T
\approx 1.4 \times 10^{-2}$ as a function of $N$, for example, gives us an
estimate of $\beta/\nu' = 0.339(2)$, which is fairly close to $1/3$. Then we
can estimate $z' \approx 0.3$ by making the data for different $N$'s
collapse into a single curve at large $b$ [see, e.g.,
Fig.~\ref{fig:global2}(a)]. One might say that the mean-field behavior could
have been found by using $\left< |m| \right>$ instead [see
Fig.~\ref{fig:global2}(b)]. But our point is that $\left< |s| \right>$ is
easier to handle in the sense that it converges a well-defined value in the
large-$b$ limit, which allows us to get around the problem of working with
two scaling parameters at the same time. Beginning with scaling $\left< |s|
\right>$, one finds numerical observables such as $U$ and $\left< |m|
\right>$ consistent with the mean-field behavior [Figs.~\ref{fig:global2}(c)
and \ref{fig:global2}(d)]. Therefore, we suggest that Eq.~(\ref{eq:sfss})
can be a practically useful guidance in measuring the
dynamic critical exponent. The advantage of using $U$, on the other hand,
comes from its universal values containing a good deal of information on
universality classes, and it indeed works well in estimating $\Gamma_c$.
Figures~\ref{fig:global2}(a) and \ref{fig:global2}(b) furthermore
suggest that a relevant scaling region should be found at $b \gg 1$, rather
than around the peak position. Once again, we note that $U$ vanishes at this
large $b$, while $\left< |s| \right>$ does not vanish for finite $N$, so
that the latter quantity can be more easily handled at the scaling region.

\begin{figure}
\includegraphics[width=0.48\textwidth]{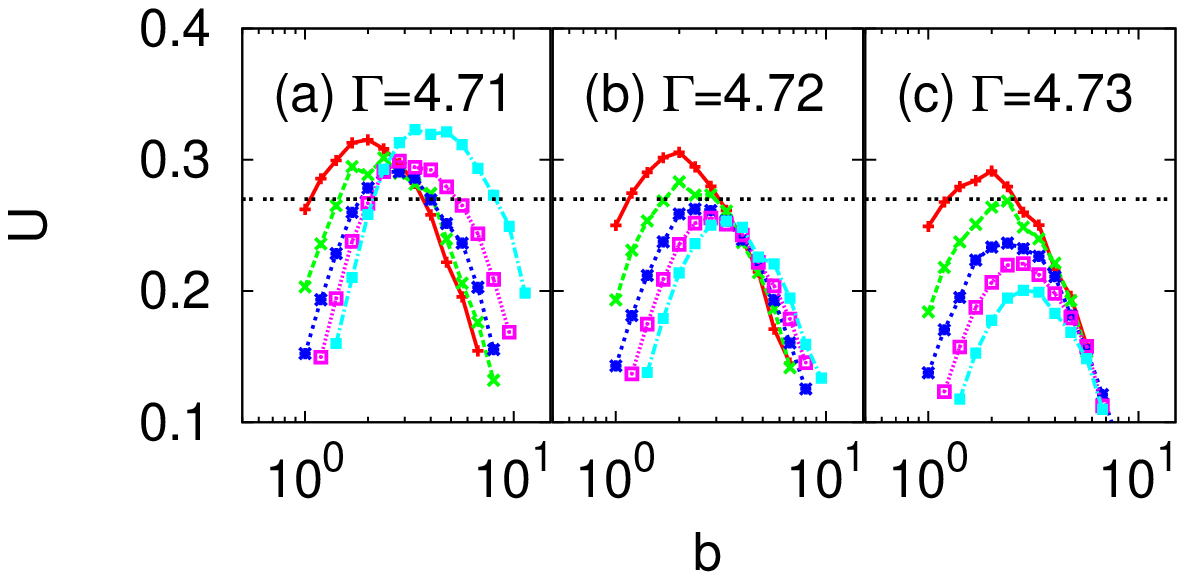}
\includegraphics[width=0.48\textwidth]{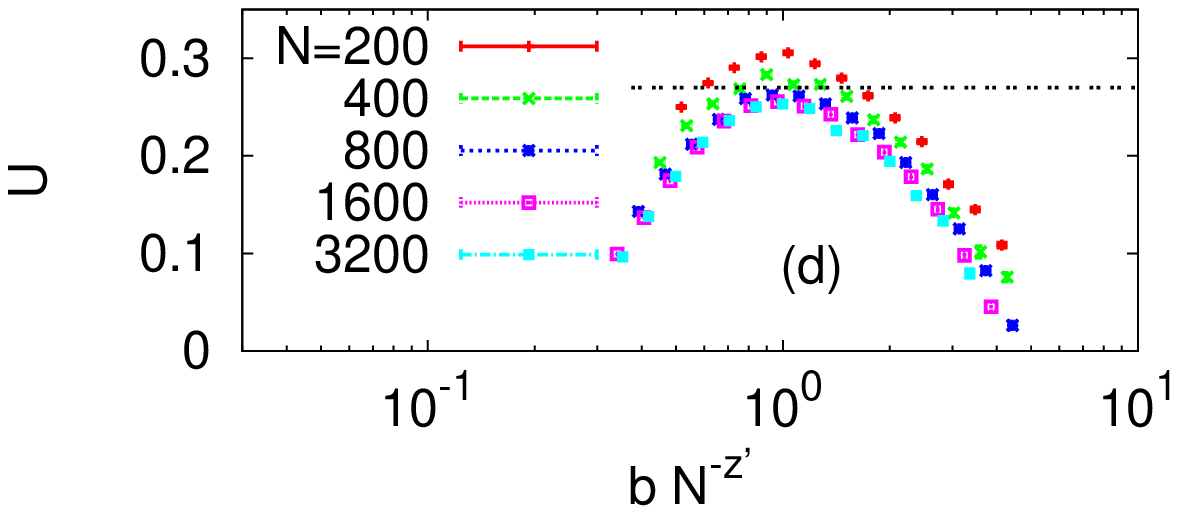}
\caption{(Color online) Observations for the WS network. (a)--(c) Binder's
cumulant as a function of $b$ at three values of $\Gamma$. In each panel,
the curves represent $N=200, 400, 800, 1600$, and $3200$ from left to right,
respectively. (d) From Eq.~(\ref{eq:ufss}), we measure the dynamic critical
exponent $z' = 0.197(8)$, whereby this scaling collapse is obtained. The
dotted lines represent the universal amplitude ratio for the mean-field case
[$U^{\ast} \approx 0.270520$]. }
\label{fig:ws}
\end{figure}

\begin{figure}
\includegraphics[width=0.48\textwidth]{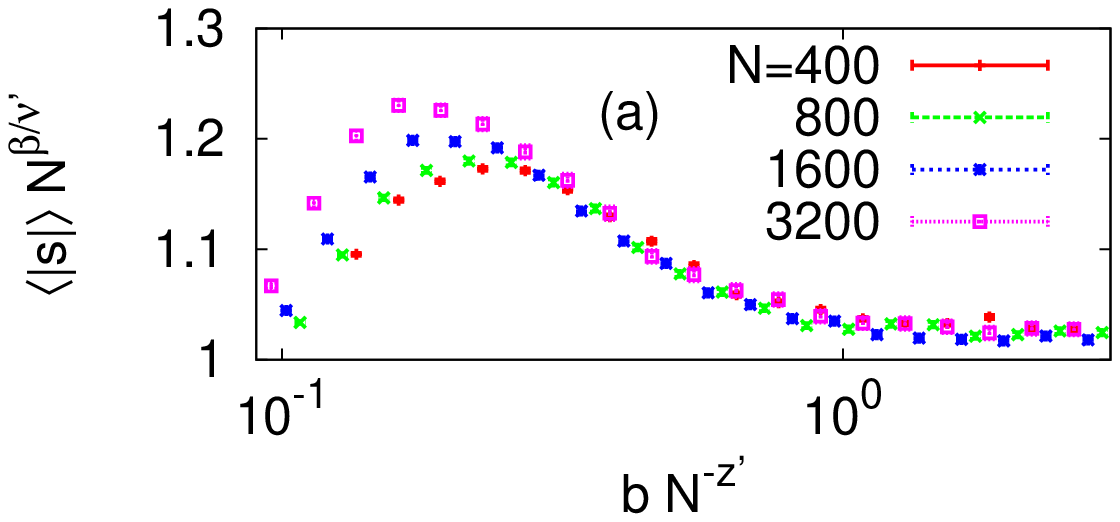}
\includegraphics[width=0.48\textwidth]{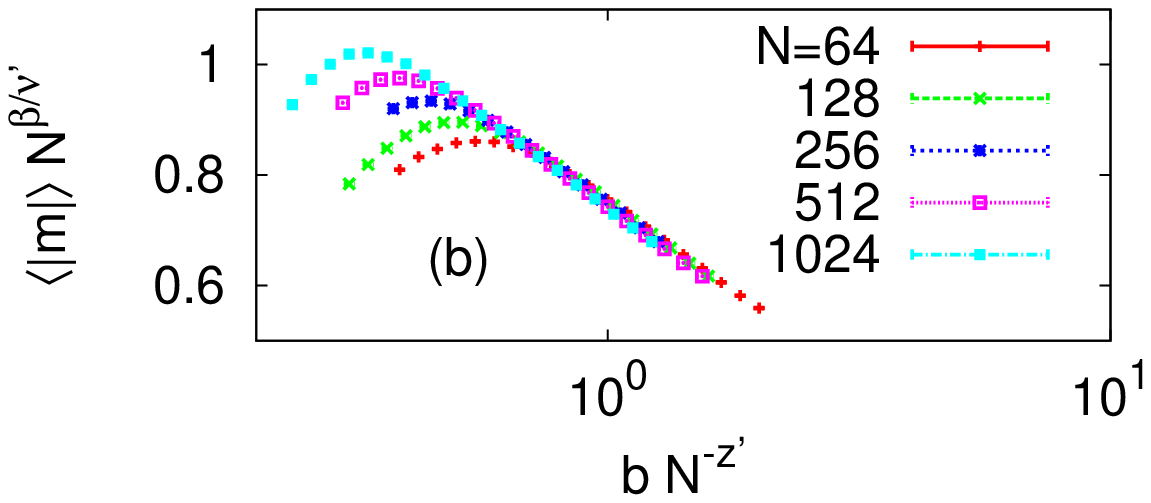}
\includegraphics[width=0.48\textwidth]{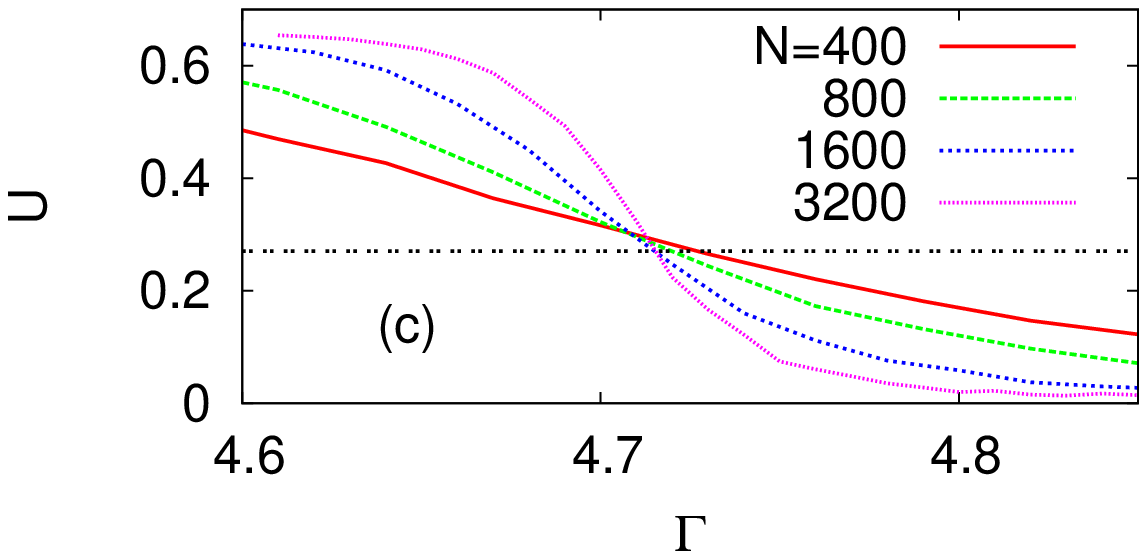}
\includegraphics[width=0.48\textwidth]{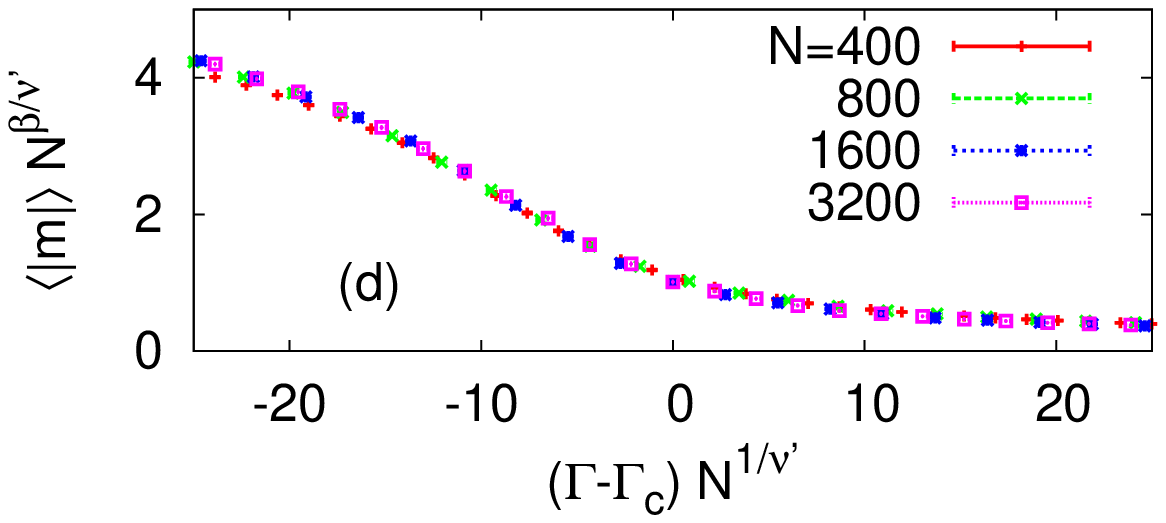}
\caption{(Color online) Observations for the WS network. (a) Scaling
collapse of $\left< |s| \right>$ at $\Gamma = \Gamma_c^{\rm (WS)} = 4.72$
according to Eq.~(\ref{eq:sfss}), with $\beta/\nu' = 1/3$ and $z' = 1/3$.
(b) Scaling collapse of the magnetic order parameter with $z'=1/3$. Choosing
this $z'$, we find (c) the crossing of $U$ and (d) scaling collapse of
$\left< |m| \right>$ at $\Gamma = 4.72$ with the mean-field values, i.e.,
$\beta/\nu' = 1/3$ and $1/\nu' = 2/3$. The dotted line indicates the
universal amplitude ratio for the mean-field case [$U^{\ast} \approx
0.270520$]. }
\label{fig:ws2}
\end{figure}

Let us now proceed to the WS network. We choose $k=3$ and $P=0.05$ and take
disorder averages over independent realizations. The number of realizations
varies but usually amounts to $O(10^3)$. We begin with the existing method
using $U$, assuming that it will work here as in 1D. Then we find
characteristic temperatures for various sizes of the WS network by measuring
$U$, from which we estimate $\Gamma_c^{\rm (WS)} = 4.72(1)$.
[Figs.~\ref{fig:ws}(a) to \ref{fig:ws}(c)]. Noting that the small sizes tend
to overestimate $U$, we determine $\Gamma_c$ by observing the largest sizes.
It is interesting that we do not see such large asymmetry as in the globally
coupled network [compare Fig.~\ref{fig:global}(a) with
Fig.~\ref{fig:ws}(d)], which might be due to the fact that the length scale
of the WS network is a logarithmic function of $N$. Another interesting
indication is that the height of $U_{\max}$ at $\Gamma = 4.72$ is close to
the mean-field universal amplitude ratio [Figs.~\ref{fig:ws}(b) and
\ref{fig:ws}(d)]. The dynamic critical exponent at $\Gamma = 4.72$ is
estimated as $z' = 0.197(8)$ from the dependence of $T_{\rm max}$ on $N$
[Fig.~\ref{fig:ws}(d)]. It is apparently very close to $1/5$, and leads to
$\beta/\nu' = 0.296(8)$ and $1/\nu' = 0.598(16)$, or, in other words, $\beta
= 0.49(3)$ and $\nu' = 1.67(4)$. Although they are close to the mean-field
values $1/2$ and $3/2$, respectively, we have no explanation for $z' =
1/5$.
 
Since we already know that Eq.~(\ref{eq:ufss}) does not always give correct
$z'$, we may instead try the alternative method relying upon
Eq.~(\ref{eq:sfss}). By fitting the values of $\left< |s| \right>$ at $T
\approx 1.9 \times 10^{-2}$ with different $N$'s, we find $\beta/\nu' =
0.33(1)$, which contains $1/3$ within the error bar. The dynamic critical
exponent is then estimated as $z' \approx 0.3$ by making the data collapse
[see, e.g., Fig.~\ref{fig:ws2}(a)]. This method definitely prefers $z' =
1/3$ to $1/5$, which implies that the use of Eq.~(\ref{eq:ufss}) might yield
a spurious estimate of $z'$ as in the globally coupled network. As seen in
Fig.~\ref{fig:ws2}(a), our observations of $\beta/\nu'$ and $z'$ are fully
consistent since both of them are explained by $d_u = 3$. Note also the nice
agreement in Fig.~\ref{fig:ws2}(b). Accepting this $z'=1/3$, the crossing
point of $U$ is also consistent with the universal amplitude ratio of the
mean-field type, and the best fits with $\left< |m| \right>$ and $dU /
d\Gamma$ give us $\beta = 0.50(3)$ and $\nu' = 1.38(4)$
[Fig.~\ref{fig:ws2}(c)]. The value of $\nu'$ is reasonably close to $3/2$
and our numerical results do not rule out this value
[Fig.~\ref{fig:ws2}(d)]. Overall, we may consider our Monte Carlo results
for the WS network as supporting the mean-field behavior, i.e., $\beta =
1/2$, $\nu' = 3/2$, and $U^{\ast} \approx 0.270520$. It is worth noting that
Eq.~(\ref{eq:sfss}) again outperforms Eq.~(\ref{eq:ufss}) in the finite-size
scaling analysis.

In summary, we have studied the transverse-field Ising model on
infinite-dimensional structures by quantum Monte Carlo simulations. Since
the world-line Monte Carlo algorithm cannot directly access $T=0$, we need
the dynamic critical exponent $z'$ for each of the structures. For the WS
network, we have estimated $z' = 0.3(1)$ and $\beta/\nu'=0.33(1)$ from
$\left< |s| \right>$, or $\beta = 0.50(3)$ and $\nu' = 1.38(4)$ from $\left<
|m| \right>$ and $dU/d\Gamma$ under the assumption that $z'=1/3$. These
strongly support the mean-field behavior with the upper critical dimension
$d_u = 3$, i.e., $\beta = 1/2$ and $\nu' = 3/2$. Drawing this conclusion, we
have also found that the existing finite-size scaling method using $U$ can
give incorrect values of $z'$ in the infinite-dimensional structures. For
this reason, we have suggested an alternative observable $\left< |s|
\right>$ based on the quantum-mechanical expectation value, which is found
to converge more quickly to the correct scaling behavior than $U$ does.

\acknowledgments
We thank Hangmo Yi for discussions and comments.
This work was supported by a National Research Foundation of Korea (NRF)
grant funded by the Korea government (MEST) (Grant No. 2011-0015731).
%B. J. K. was supported by a Korea Research Foundation Grant
%KRF-2009-013-C00021.
This research was conducted using the
resources of the High Performance Computing Center North (HPC2N).
We thank R. Or\'{u}s and J. Vidal for informing us of relevant works on
the Lipkin-Meshkov-Glick model.~\cite{dusuel}

%\bibliographystyle{revtex}
%\bibliography{qi}

\begin{thebibliography}{10}
\providecommand*{\bibinfo}[2]{#2}
\providecommand*{\eprint}[1]{#1}
\providecommand*{\url}[1]{#1}
\bibitem{doro}
\bibinfo{author}{S.~N. Dorogovtsev}, \bibinfo{author}{A.~V. Goltsev}, and
  \bibinfo{author}{J.~F.~F. Mendes}, \bibinfo{journal}{Rev. Mod. Phys.}
  \bibinfo{volume}{\textbf{80}}, \bibinfo{pages}{1275} (\bibinfo{date}{2008}).
\bibitem{ws}
\bibinfo{author}{D.~J. Watts} and \bibinfo{author}{S.~H. Strogatz},
  \bibinfo{journal}{Nature (London)} \bibinfo{volume}{\textbf{393}},
  \bibinfo{pages}{440} (\bibinfo{date}{1998}).
\bibitem{moore}
\bibinfo{author}{C.~Moore} and \bibinfo{author}{M.~E.~J. Newman},
  \bibinfo{journal}{Phys. Rev. E} \bibinfo{volume}{\textbf{62}},
  \bibinfo{pages}{7059} (\bibinfo{date}{2000}).
\bibitem{smallxy}
\bibinfo{author}{B.~J. Kim}, \bibinfo{author}{H.~Hong},
  \bibinfo{author}{P.~Holme}, \bibinfo{author}{G.~S. Jeon},
  \bibinfo{author}{P.~Minnhagen}, and \bibinfo{author}{M.~Y. Choi},
  \bibinfo{journal}{Phys. Rev. E} \bibinfo{volume}{\textbf{64}},
  \bibinfo{pages}{056135} (\bibinfo{date}{2001}).
\bibitem{hong}
\bibinfo{author}{H.~Hong}, \bibinfo{author}{B.~J. Kim}, and
  \bibinfo{author}{M.~Y. Choi}, \bibinfo{journal}{Phys. Rev. E}
  \bibinfo{volume}{\textbf{66}}, \bibinfo{pages}{018101}
  (\bibinfo{date}{2002}).
\bibitem{med}
\bibinfo{author}{K.~Medvedyeva}, \bibinfo{author}{P.~Holme},
  \bibinfo{author}{P.~Minnhagen}, and \bibinfo{author}{B.~J. Kim},
  \bibinfo{journal}{Phys. Rev. E} \bibinfo{volume}{\textbf{67}},
  \bibinfo{pages}{036118} (\bibinfo{date}{2003}).
\bibitem{ref}
\bibinfo{author}{H.~Hong}, \bibinfo{author}{M.~Ha}, and
  \bibinfo{author}{H.~Park}, \bibinfo{journal}{Phys. Rev. Lett.}
  \bibinfo{volume}{\textbf{98}}, \bibinfo{pages}{258701}
  (\bibinfo{date}{2007}).
\bibitem{nagaj}
\bibinfo{author}{D.~Nagaj}, \bibinfo{author}{E.~Farhi},
  \bibinfo{author}{J.~Goldstone}, \bibinfo{author}{P.~Shor}, and
  \bibinfo{author}{I.~Sylvester}, \bibinfo{journal}{Phys. Rev. B}
  \bibinfo{volume}{\textbf{77}}, \bibinfo{pages}{214431}
  (\bibinfo{date}{2008}).
\bibitem{krz}
\bibinfo{author}{F.~Krzakala}, \bibinfo{author}{A.~Rosso},
  \bibinfo{author}{G.~Semerjian}, and \bibinfo{author}{F.~Zamponi},
  \bibinfo{journal}{Phys. Rev. B} \bibinfo{volume}{\textbf{78}},
  \bibinfo{pages}{134428} (\bibinfo{date}{2008}).
\bibitem{yi}
\bibinfo{author}{H.~Yi} and \bibinfo{author}{M.-S. Choi},
  \bibinfo{journal}{Phys. Rev. E} \bibinfo{volume}{\textbf{67}},
  \bibinfo{pages}{056125} (\bibinfo{date}{2003}).
\bibitem{yi3}
\bibinfo{author}{H.~Yi}, \bibinfo{journal}{Eur. Phys. J. B}
  \bibinfo{volume}{\textbf{61}}, \bibinfo{pages}{89} (\bibinfo{date}{2008}).
\bibitem{yi2}
\bibinfo{author}{H.~Yi}, \bibinfo{journal}{Phys. Rev. E}
  \bibinfo{volume}{\textbf{81}}, \bibinfo{pages}{012103}
  (\bibinfo{date}{2010}).
\bibitem{elliott}
\bibinfo{author}{R.~J. Elliott}, \bibinfo{author}{P.~Pfeuty}, and
  \bibinfo{author}{C.~Wood}, \bibinfo{journal}{Phys. Rev. Lett.}
  \bibinfo{volume}{\textbf{25}}, \bibinfo{pages}{443} (\bibinfo{date}{1970}).
\bibitem{pfeuty}
\bibinfo{author}{P.~Pfeuty}, \bibinfo{journal}{J. Phys. C}
  \bibinfo{volume}{\textbf{9}}, \bibinfo{pages}{3993} (\bibinfo{date}{1976}).
\bibitem{um}
\bibinfo{author}{J.~Um}, \bibinfo{author}{S.-I. Lee}, and
  \bibinfo{author}{B.~J. Kim}, \bibinfo{journal}{J. Korean Phys. Soc.}
  \bibinfo{volume}{\textbf{50}}, \bibinfo{pages}{285} (\bibinfo{date}{2007}).
\bibitem{botet}
\bibinfo{author}{R.~Botet} and \bibinfo{author}{R.~Jullien},
  \bibinfo{journal}{Phys. Rev. B} \bibinfo{volume}{\textbf{28}},
  \bibinfo{pages}{3955} (\bibinfo{date}{1983}).
\bibitem{kawa2}
\bibinfo{author}{N.~Kawashima} and \bibinfo{author}{K.~Harada},
  \bibinfo{journal}{J. Phys. Soc. Jpn} \bibinfo{volume}{\textbf{73}},
  \bibinfo{pages}{1379} (\bibinfo{date}{2004}).
\bibitem{wolff}
\bibinfo{author}{U.~Wolff}, \bibinfo{journal}{Phys. Rev. Lett.}
  \bibinfo{volume}{\textbf{62}}, \bibinfo{pages}{361} (\bibinfo{date}{1989}).
\bibitem{kawa}
\bibinfo{author}{H.~Rieger} and \bibinfo{author}{N.~Kawashima},
  \bibinfo{journal}{Eur. Phys. J. B} \bibinfo{volume}{\textbf{9}},
  \bibinfo{pages}{233} (\bibinfo{date}{1999}).
\bibitem{sachdev}
\bibinfo{author}{S.~Sachdev}, \bibinfo{title}{\emph{Quantum phase transitions}}
  (\bibinfo{publisher}{Cambridge University Press}, Cambridge,
  \bibinfo{year}{1999}).
\bibitem{guo}
\bibinfo{author}{M.~Guo}, \bibinfo{author}{R.~N. Bhatt}, and
  \bibinfo{author}{D.~A. Huse}, \bibinfo{journal}{Phys. Rev. Lett.}
  \bibinfo{volume}{\textbf{72}}, \bibinfo{pages}{4137} (\bibinfo{date}{1994}).
\bibitem{young}
\bibinfo{author}{H.~Rieger} and \bibinfo{author}{A.~P. Young},
  \bibinfo{journal}{Phys. Rev. Lett.} \bibinfo{volume}{\textbf{72}},
  \bibinfo{pages}{4141} (\bibinfo{date}{1994}).
\bibitem{salas}
\bibinfo{author}{J.~Salas} and \bibinfo{author}{A.~D. Sokal},
  \bibinfo{journal}{J. Stat. Phys.} \bibinfo{volume}{\textbf{98}},
  \bibinfo{pages}{551} (\bibinfo{date}{2000}).
\bibitem{lb}
\bibinfo{author}{E.~Luijten} and \bibinfo{author}{H.~W.~J. Bl{\"o}te},
  \bibinfo{journal}{Int. J. Mod. Phys. C} \bibinfo{volume}{\textbf{6}},
  \bibinfo{pages}{359} (\bibinfo{date}{1995}).
\bibitem{cross}
\bibinfo{author}{W.~A.~C. Erkelens}, \bibinfo{author}{L.~P. Regnault},
  \bibinfo{author}{J.~Rossat-Mignod}, \bibinfo{author}{J.~E. Moore},
  \bibinfo{author}{R.~A. Butera}, and \bibinfo{author}{L.~J. de~Jongh},
  \bibinfo{journal}{EPL} \bibinfo{volume}{\textbf{1}}, \bibinfo{pages}{37}
  (\bibinfo{date}{1986}).
\bibitem{dusuel}
\bibinfo{author}{S.~Dusuel} and \bibinfo{author}{J.~Vidal},
  \bibinfo{journal}{Phys. Rev. Lett.} \bibinfo{volume}{\textbf{93}},
  \bibinfo{pages}{237204} (\bibinfo{date}{2004});
\bibinfo{author}{P.~Ribeiro}, \bibinfo{author}{J.~Vidal}, and
  \bibinfo{author}{R.~Mosseri}, \bibinfo{journal}{{\it ibid.}}
  \bibinfo{volume}{\textbf{99}}, \bibinfo{pages}{050402}
  (\bibinfo{date}{2007});
\bibinfo{author}{R.~Or\'{u}s}, \bibinfo{author}{S.~Dusuel}, and
  \bibinfo{author}{J.~Vidal}, \bibinfo{journal}{{\it ibid.}}
  \bibinfo{volume}{\textbf{101}}, \bibinfo{pages}{025701}
  (\bibinfo{date}{2008}).

\end{thebibliography}

\end{document}